\documentclass[preprint,aps,showpacs]{revtex4}

\usepackage{graphicx}

\begin{document}

\title{Introduction to Topological Insulators}

\author{Hongming Weng, Xi Dai, Zhong Fang}

\affiliation{Beijing National Laboratory for Condensed Matter Physics,
  and Institute of Physics, Chinese Academy of Sciences, Beijing
  100190, China;}

\date{Jul. 5, 2011}

\begin{abstract}
  In this article, we will give a brief introduction to the
  topological insulators. We will briefly review
  some of the recent progresses, from both theoretical and
  experimental sides. In particular, we will emphasize the recent
  progresses achieved in China. 
\end{abstract}

\maketitle

\section{Introduction.}

Topological insulator (TI) has drawn extensive attentions recently in
the field of condensed matter physics, not only because of its
fundamental importance but also because of its potential applications
for future technology. One of the most simple and distinct properties
of TIs is that it is electronically insulating in bulk while
conducting along its boundary (for examples, edges in two dimensional
systems and surfaces in three dimensional systems), due to the
topologically unavoidable gapless boundary states. The existence of
metallic boundary states is very robust, and they are protected by
topological invariants. Though this property is in sharp contrast to
our general knowledge on normal insulators or simple metals, similar
phenomena has been observed and well-known since 1980, due to the
discovery of quantum Hall effect (QHE)~\cite{qhe}. In QHE, electrons
in two-dimensional (2D) materials are enforced to change its quantum
state into a new one, so called Landau energy level, under highly
intensive magnetic field. The original free-electron-like conducting
electrons start to make cyclotron motion. It is easy to understand
that in the bulk of the material, full circle cyclotron motion leads
to electron localization, resulting in an insulating bulk. While along
the edge of the 2D system, the circle motion of electrons enforced by
the magnetic field can not be completed due to the presence of edge,
which make the electrons travel in a way forming so called edge state,
and electrons in such state can circumambulate defects or impurities
on their way ``smartly" (due to the absence of back scatterings).
Therefore, current carried by these electrons is dissipationless and
conductance is quantized into unit of $e^2/\hbar$ with quantum number
corresponding to the number of edge states. Such fascinating quantum
state and physical phenomena are highly interesting and impact the
whole field of physics. Since then, people started to realize that this
is a new state of quantum matters and it should be characterized by
topology of electronic wavefunctions. This topological number can be
evaluated from the formula given by D. J. Thouless, M. Kohmoto, M. P.
Nightingale and M. den Nijs~\cite{tknn}. It is called as TKNN number
or first Chern number. This topological number has direct physical
meaning and is experimentally observable, because it is the same as
the number of edge states or the quantum number of Hall conductance.


Obviously the application of QHE requires highly intensive magnetic
field and low temperature. In 1988, F. D. M. Haldane~\cite{haldane}
proposed a genius model to realize the quantum Hall state in solid
state materials without introducing Landau levels. His model is based
on graphene like honeycomb lattice structure, which has two
sublattices. If equal and opposite magnetic fluxes can be threaded
through these two sublattices, respectively, the electrons hopping at
least upto next nearest neighbor within such lattice model would be in
a quantum state similar to the quantum Hall state. While the possible
materials realizing Haldane's dream is still under intensive pursuing
upto today, on the other hand, in 2005, C. L. Kane and E. J.
Mele~\cite{kanemele} made another big step forward to reach Haldane's
dream. They demonstrated that spin-orbit coupling, which is inherent
in any material, can effectively play a role of the imagined magnetic
field in Haldane's lattice model. Different from the quantum Hall
state, however, the time reversal symmetry is preserved in the later
case. As C. L. Kane and E. J. Mele pointed out, the electrons within
such graphene model are in another new quantum state, characterized by
a topological invariant Z$_2$ number, which is also determined by the
topology of wavefunctions describing this state. Electrons in this state 
behave in the similar way as in QHE, namely they are 
insulating in the bulk but conducting on the edges. Insulator possessing 
this kind of quantum state is coined as ``topological insulator" by J. Moore 
and L. Balents~\cite{tiname}. From the non-interacting single particle
picture, in the bulk of TIs, the occupied electron-bands are well
separated from the unoccupied electron-bands by a energy gap at Fermi
level. However, due to the bands ``twisting'', the occupied states
are topologically non-trivial, and contribute to a non-zero Z$_2$
integer number. If there are edges or interfaces formed between two
insulators with different Z$_2$ number, there must appear conducting
edge states. The insulators which has ``twisted" bands are therefore
very much different from normal insulator or vacuum with ``untwisted"
bands. Compared with QHE, 2D TI has at least two edge states for 
each edge, due to the time reversal symmetry. The
electrons in two edge stats must have opposite spins, and they must
travel in opposite directions. This is exactly the quantized spin Hall
effect~\cite{QSHE}, therefore, 2D TI is also called
as quantum spin Hall insulator.

Though graphene is predicted as a prototype 2D TI,
Y. Yao {\it et al.} found that the spin-orbit coupling in graphene is
negligibly small~\cite{yao} and pointed out that the bulk insulating
gap is too small to observe quantum spin Hall effect. In 2006, B. A.
Bernevig, T. A. Hughes and S. C. Zhang~\cite{hgte} theoretically
predicted that the HgTe quantum well can be tuned into topological
insulating phase and its edge states can give out quantum spin Hall
effect. They also established an intuitive model hamiltonian to
describe the topological phase transition driven by ``band inversion".
Soon in 2007, quantum spin Hall effect shows up in HgTe quantum well
insulator just as predicted in the experiment done by M. K\"{o}nig
{\it et al.}~\cite{hgteexp}. And a detailed tight-binding model study
on HgTe is given by X. Dai {\it et al.}~\cite{hgtetb}. Recently, C. C.
Liu {\it et al.} proposed that Silicene~\cite{Silicene} with
low-buckled sheet structure with enhanced spin-orbit coupling might
show detectable quantum spin Hall effect, which might be suitable for
applications in spintronic devices since it is compatible with current
silicon-based microelectronic industry.

\section{Searching for three-dimensional Topological Insulators.}

Nearly at the same time in 2006, L. Fu {\it et al.}~\cite{3dti}, J.
Moore {\it et al.}~\cite{tiname}, as well as R. Roy~\cite{3dti2},
extended the concept of TIs to three-dimensional (3D) case. Similar as
2D HgTe quantum well, 3D Bi$_{1-x}$Sb$_x$ alloy is predicted~\cite{3dti,
  murakami,BiSb-HJ} by theoretical calculations to have topological
nontrivial conducting surface state depending on $x$ value and then it
is confirmed by experimental measurements~\cite{bisb}, though the
topologically protected surface state is complex and the bulk band gap
is too small for possible applications. The important realization of
3D TIs are tetradymite semiconductors Bi$_2$Se$_3$,
Bi$_2$Te$_3$ and Sb$_2$Te$_3$, which are, again, predicted by
theoretical calculations~\cite{Zhang} and confirmed by experimental
measurements~\cite{Xia, Chen}. This new family 3D TIs 
have single Dirac cone like surface states and bulk band
gap is as larger as 0.3 eV, which simplifies the further study on the
properties of surface states and promises the potential applications
even under room temperature~\cite{nextg}. After this, lots of efforts
have been performed to find various 3D TIs. B. Yan
{\it et al.}~\cite{yan2010} and H. Lin {\it et al.}~\cite{lin} have
independently predicted that thallium-based III-V-VI$_2$ ternary
chalcogenides might be 3D TIs, and later
experimental works done by T. Sato {\it et al.}~\cite{tlbise1} and K.
Kuroda {\it et al.}~\cite{tlbise2} have confirmed their predictions.
Some other ternary half-Heusler compounds~\cite{half-heusler1,
  half-heusler2, xiao}, chalcopyrite compounds~\cite{feng} and
materials with antiperovskite structure~\cite{antiperovskite}, as well
as honeycomb lattice~\cite{honey}, have been predicted to be 3D TI
with or without additional uniaxial strain. Recently, another new
binary TI Ag$_2$Te has been proposed by W. Zhang
{\it et al.}~\cite{Ag2Te}, which has highly anisotropic Dirac cone
surface state, in contrast to known examples. Therefore, there are
plenty of materials having nontrivial topological state. They supply a
solid material basis for studying various intriguing physical
properties of surface states, as well as potential applications.
Theoretical predictions based on first-principles calculations have
played a crucial important role in searching for new TIs, 
which accelerates the study of this field tremendously.


The defining property of TIs is the topologically
protected surface state, which is spin-momentum locked, and $\pi$
Berry phase is acquired for the wavefunctions circling the Dirac cone.
Angle-resolved photoemission spectroscopy (ARPES) is the usual and
widely used method to directly observe this surface state~\cite{bisb,
  Xia, Chen, tlbise1, tlbise2}. To detect the chiral spin texture of
surface state, as well as $\pi$ Berry phase, one useful technique is
the spin-resolved ARPES, which is firstly applied on Bi$_{1-x}$Sb$_x$
alloy~\cite{spinarpes} and then on Bi$_2$Se$_3$~\cite{spinarpesbise},
Bi$_2$Te$_3$~\cite{spinarpesbite,
  spinarpesbite2} and TlBiSe$_2$~\cite{spinarpesbite}. When surface
states are close to the bulk states, they start to deform from the
circle-like Dirac cone and follow the bulk crystal symmetry. The spin
texture will be modified at the same time. Such effect is demonstrated
by ARPES or spin-resolved ARPES for Bi$_2$Te$_3$~\cite{Chen,
  spinarpesbite, spinarpesbite2, adv1}, where Fermi level is
systematically tuned by carrier doping. This effect has been explained
well by an effective model~\cite{warp} and reproduced by
first-principles calculations~\cite{zhangbisb, Ag2Te, Wei}.


The spin-momentum locked conducting surface state is predicted to have
one important feature: the conducting electron will not be
back-scattered or localized by non-magnetic (time reversal symmetric)
scattering potential. One evidence of this feature is the measurement
of interference patterns on the surface of TI by
scanning tunneling spectroscopy. T. Zhang {\it et
  al.}~\cite{backscattering1} confirm that the backscattering of
surface states is absent on non-magnetic Ag doped Bi$_2$Te$_3$ (111)
surface. P. Roushan {\it et al.}~\cite{backscattering2} show that the
atomic disorder in Bi$_{1-x}$Sb$_{x}$ will not cause backscattering of
surface states. Y. Okada {\it et al.}~\cite{backscattering3}
demonstrate that magnetic impurity Fe doped Bi$_2$Te$_3$ breaks time
reversal symmetry and the backscattering does exist. Nevertheless, the
direct transport measurements of surface conducting states are
hindered since the residual conduction in the bulk from impurities or
defects in sample can easily overcome the surface contributions. H.
Peng {\it et al.}~\cite{abeffect} circumvent this problem by preparing
Bi$_2$Te$_3$ nanoribbon, which has much large surface/volume ratio.
They observed Aharonov-Bohm interference for the currents conducting
through the surface of TI nanoribbon. J. Chen {\it
  et. al.}~\cite{walbise} successfully grow Bi$_2$Se$_3$ on SrTiO$_3$
substrates, which enables very large tunablity in carrier density with
a back gate. They observed weak antilocalization (WAL) effect in
magnetoresistance measurements, which originates from the $\pi$ Berry
phase obtained by surface electrons circling the Dirac cone. H.-T. He
{\it et al.}~\cite{walbite} further demonstrated that the non-magnetic
impurity will not influence the WAL effect while the magnetic impurity
quenches the WAL effect in Bi$_2$Te$_3$. M. Liu {\it et
  al.}~\cite{wal2wlcrBiSe} have done the similar experiments on Cr
doped Bi$_2$Se$_3$. The direct evidence that surface states can
overcome the non-magnetic lattice imperfection is the experiment done
by J. Seo {\it et al.}~\cite{sbtrans} on Sb surface, where they
observed that the topological surface states can penetrate the atomic
steps instead of being reflected or absorbed as in common metals.

\section{Opening Gap on Surface States: towards Applications}

Many of the potential applications of TIs are
related to the gap opening of topological surface states. Some of the
possibilities will be discussed in this section. Since time reversal
symmetry is conserved in 3D TI, there should be two Dirac cones
according to fermion doubling theorem~\cite{fermiondouble}. They are
in fact located on the opposite surfaces, which are well separated by
the insulating bulk. W. Zhang {\it et
  al.}~\cite{Wei} has shown that in Bi$_2$Se$_3$ the surface states
can penetrate into bulk by about 4 nm. Therefore, if film sample is
thin enough, the two surface states can have finite coupling and a gap
would open at Dirac cone point. C.-X. Liu {\it et
  al.}~\cite{thinfilmliu} have pointed out that there will be an
oscillatory crossover from 2D to 3D
TI as film thickness increases. This crossover
behavior has been observed by ARPES experiments performed on
Bi$_2$Se$_3$~\cite{thinfilmBi2Se3,thinfilmBi2Se3sec} and
Bi$_2$Te$_3$~\cite{thinfilmBi2Te3}.


In addition to the coupling between two surface states, magnetic
impurities which break time reversal symmetry can also open a gap at
the Dirac point of surface states. Fe ions have been introduced into
Bi$_2$Se$_3$~\cite{biseFe}, and ARPES experiment has observed the
opening of gap. Mn and Fe ions have also been introduced into
Bi$_2$Te$_3$~\cite{biteMn, biteFe} and similar ARPES experiments have
confirmed the gap opening. R. Yu {\it et al.}~\cite{qahe} have
proposed that ferromagnetic TI can be obtained in
Cr and Fe doped thin film Bi$_2$Se$_3$ and quantum anomalous Hall effect
can be realized in this way. When perpendicular external magnetic
field is applied onto the surface state, Landau level will appear. The
Landau level for Dirac electrons is different from that for the
parabolic free electrons. There is one zero energy Landau level
independent of magnetic field strength $\bf B$. Other Landau level
energy is proportional to $\sqrt{B}$~\cite{landaudirac}. P. Chen {\it
 et al.}~\cite{landau1} and T. Hanaguri {\it et al.}~\cite{landau1}
have observed such Landau quantization of surface states in
Bi$_2$Se$_3$ independently by scanning tunneling microscopy and
spectroscopy. This kind of Landau level is also the basis for theory
of quantum magnetoresistance proposed by A. A.
Abrikosov~\cite{quantummag}, which is used to explain the observed
non-saturated linear magnetoresistance in non-magnetic
Ag$_2$Te~\cite{Ag2Te}.


Introducing time-reversal-breaking effects into TI
can lead to topological magnetoelectric effect. The electromagnetic
response of magnetic TI is the same as axion
electrodynamics. Therefore, the search for and study of axion
insulator becomes another important direction in sense of finding
large magnetoeletric coupling materials. Na$_2$IrO$_3$~\cite{NaIrO} is
firstly proposed to be a TI, which contains 5$d$
transition metal possessing competing electron-electron correlation
interaction $U$ and spin-orbit coupling. D. Pesin and L.
Balents~\cite{pyrochlore} then suggest that certain iridium based
pyrochlore compounds such as Pr$_2$Ir$_2$O$_7$ may be 3D TIs. 
X. Wan {\it et al.}~\cite{wan1} further studied pyrochlore
Y$_2$Ir$_2$O$_7$ by first-principles calculations. They found rich
phase diagram of Y$_2$Ir$_2$O$_7$ as electron-electron correlation
interaction $U$ changes, including the topologically nontrivial axion
insulator, Weyl semimetal, etc. The Weyl semimetal phase has
intriguing fermi arc as its surface states, although this phase is
sensitive to the parameter $U$ used in their calculations and might
not be the nature state of the material. X. Wan {\it et
  al.}~\cite{wan2} further predicted that hypothetical Osmium spinel
compounds such as CaOs$_2$O$_4$ and SrOs$_2$O$_4$ might have the
similar phase diagram depending on $U$. Recently, G. Xu {\it et
  al.}~\cite{xu} find that spinel HgCr$_2$Se$_4$ is a Weyl semimetal
in its natural ferromagnetic ground state. The HgCr$_2$Se$_4$ is in
fact a so called ``Chern semi-metal'', because the Weyl fermions are
realized at the topological phase boundary separating the different
insulating layers (in momentum $\vec{k}$-space) with different Chern
numbers, where the Weyl fermions are topologically unavoidable. It is
further proposed that the quantum anomalous Hall effect is a direct
observable effect in quantum well of Hg$_2$CrSe$_4$, which might be
confirmed by future experiments.


Besides these, breaking gauge symmetry due to proximity to a
superconductor can also open a gap on topological surface states. L.
Fu and C. L. Kane~\cite{proximity} proposed that the proximity effect
between an $s$-wave superconductor and the surface states will result
in a 2D state resembling a spinless $p_x+ip_y$ superconductor without
breaking time reversal symmetry. This state supports Majorana bound
states at vortices. Majorana fermions are particles which are their
own antiparticles~\cite{majorana}. They constitute only half of a
usual fermion, and obey the non-Abelian statistics~\cite{non-Abelian},
which is the key ingredient for the fault-tolerant topological quantum
computation~\cite{TQC}. The experimental setting can be in principles
obtained from laboratory, however, since most known 3D TIs 
are not good bulk insulators and important surface states
may overlap with bulk states, experiments have to wait for development
of well-controlled clean samples. On the other hand, a well matched
interface between TI and $s$-wave superconductor is
required to get large proximity effect. The discovery of high pressure
induced superconductivity in TI
Bi$_2$Te$_3$~\cite{highpressbite, highpressbite2} has given some hope
in realizing topological superconductivity in one compound to overcome
this problem. It was also proposed that semiconductor quantum wells
with Rashba type spin-orbit coupling in proximity to $s$-wave
superconductor will produce similar effect~\cite{Sarma, Alicea, Lee2}.
This may lower the experimental threshold, since well-controlled
samples are available nowadays. These proposals are encouraging, while
experimental obstacles still remain. First, magnetic insulating layers
or strong external magnetic field are required to break the time
reversal symmetry, which is not easy to implement experimentally;
second, the Fermi surfaces in both cases are too small, and fine
control of chemical potential is difficult for semiconductors in
contact with a superconductor. Using half-metal with odd number Fermi
surfaces instead of TI or semi-conductor quantum
well is then proposed~\cite{Lee1, chung, duckhein} to overcome the
above difficulties. The asymmetric interface between half-metal and
$s$-wave superconductor induces Rashba type spin-orbit coupling as the
case using semiconductor quantum well. H. Weng {\it et
 al.}~\cite{nacoo2} have proposed that NaCoO$_2$ is the right
material to realize this. They find that NaCoO$_2$ itself is
insulating while its surface is half-metalic and has just one single
large Fermi surface. The exchange splitting is as large as 0.2 eV,
which simplifies the tuning of chemical potential. These make it a
promising candidate for experimental setting up to find Majorana
fermion.

\end{document}